\documentclass[11pt]{article}
\usepackage{geometry}                
\usepackage{ifthen,color,xypic,amssymb}
\usepackage{graphicx}
\usepackage{amsmath}
\usepackage{latexsym}
\newcommand\op[1]{\operatorname{#1}}

\DeclareMathOperator{\Tr}{Tr}
\DeclareMathOperator{\tr}{tr}

\DeclareMathOperator{\Ext}{Ext}
\DeclareMathOperator{\Hom}{Hom}
\DeclareMathOperator{\td}{td}
\DeclareMathOperator{\ch}{ch}

\newcommand\iso{\kern.35em{\raise3pt\hbox{$\sim$}\kern-1.1em\to}\kern.3em}

\newcommand\fp{\times_{ B}}
\newcommand\rest[2]{{#1}_{\vert #2}}

\newcommand\Fc{{\mathcal F}}
\newcommand\Oc{{\mathcal O}}
\newcommand\Pc{{\mathcal P}}
\newcommand\M{{\mathcal M}}

\newcommand\Z{{\mathbb Z}}

\newcommand\bbQ{{\mathbb Q}}

\newcommand\Nc{{\mathcal N}}

\newcommand\Ec{{\mathcal E}}

\newcommand\Ps{{\mathbb P}}
\newcommand\X{{\widehat X}}
\newcommand\G{{\mathcal G}}

\newcommand\bS{{\Phi}}

\newcommand\FM{Fourier-Mukai transform}
\newcommand\gif{geometric integral functor}

\newcommand\FMF{Fourier-Mukai functor}

\numberwithin{equation}{section}

\title{\bf The Fourier-Mukai Transform in String Theory} 
\author{Bj\"orn Andreas\footnote{andreas@math.hu-berlin.de}\\
\it \small Institut f\"ur Mathematik\\
\it \small Humboldt-Universit\"at zu Berlin\\ }
\date{}                                           

\begin{document}

\maketitle
\begin{abstract}
The article surveys aspects of the Fourier-Mukai transform, its relative version
and some of its applications in string theory. To appear in Encyclopedia of Mathematical Physics, published by  Elsevier in early 2006. Comments/corrections welcome.
\end{abstract}

\section{Introduction}

The Fourier-Mukai transform has been introduced in the study of abelian varieties by Mukai and can be thought of as a nontrivial algebro-geometric analogue of the Fourier transform. Since its original introduction, the Fourier-Mukai transform turned out to be a useful tool for studying various aspects of sheaves on varieties and their moduli spaces, and as a natural consequence, to learn about the varieties themselves. Various links between geometry and derived categories have been uncovered, for instance, Bondal and Orlov proved that Fano varieties, and varieties of general type, can be reconstructed from their derived categories. Moreover, Orlov proved a derived version of the Torelli theorem for K3 surfaces and also a structure theorem for derived categories of abelian varieties. Later, Kawamata gave evidence to the conjecture that two birational smooth projective varieties with trivial canonical sheaves have equivalent derived categories which has been proved by Bridgeland in dimension three. 

The Fourier-Mukai transform also enters into string theory. The most prominent example is Kontsevich's homological mirror symmetry conjecture. The conjecture predicts (for mirror dual pairs of Calabi-Yau manifolds) an equivalence between the bounded derived category of coherent sheaves and the Fukaya category. The conjecture implies a correspondence between self-equivalences of the derived category and certain symplectic self-equivalences of the mirror manifold.

Besides their importance for geometrical aspects of mirror symmetry, the Fourier-Mukai transform 
has been also important for heterotic string compactifications. The motivation for this came from the conjectured correspondence between the heterotic string and F-theory which both rely on elliptically fibered Calabi-Yau manifolds. To give evidence to this correspondence, an explicit description of stable holomorphic vector bundles was necessary and inspired a series of publications by Friedman, Morgan and Witten. Their bundle construction relies on two geometrical objetcs: a hypersurface in the Calabi-Yau manifold together with a line bundle on it, more precisley, they construct vector bundles using a relative \FM. 

Various aspects and refinements of this construction have been studied by now. For instance, a physical way to understand the bundle construction can be given using the fact that holomorphic vector bundles can be viewed as D-branes and that D-branes can be mapped under T-duality to new D-branes (of different dimensions). 

We survey aspects of the Fourier-Mukai transform, its relative version
and outline the bundle construction of Friedman, Morgan and Witten. The construction has led to many new insides, for instance, the presence of five-branes in heterotic string vacua has been understood. The construction also inspired tremendous work towards a heterotic string phenomenology on elliptic Calabi-Yau manifolds. For the many topics omitted the reader should consult the references.

\section{The Fourier-Mukai transform}

Every object $E$ of the derived category on the product $X\times Y$ of two smooth algebraic 
varieties $X$ and $Y$ gives rise to a (geometric integral) functor $\bS^E$ from the bounded derived category $D(X)$  of coherent sheaves on $X$ to the similar category on $Y$
$$
\bS^{E}\colon D(X)\to D(Y) \,,\quad F\mapsto \bS^{E}(F)=
R\hat\pi_{\ast}(L\pi^\ast F\otimes E),
$$ 
where $\pi$, $\hat\pi$ are the projections from $X\times Y$ to $X$ and $Y$, respectively. $E$ and $F$ are objects of $D(X\times Y)$ and $D(Y)$ and $L\pi^\ast\colon D(X)\to D(X\times Y)$ is the derived inverse image functor. Also $E$ is usually called the ``kernel'' in analogy to the definition of an integral transfrom with kernel. Furthermore, $\otimes$ means the derived tensor product.

Now those functors that are {equivalences of categories} between $D(X)$ and $D(Y)$ are called {\FMF s} and if the kernel reduces to a single sheaf then the corresponding Fourier-Mukai functors are called {\FM s}.

In analogy to the Fourier transform there is a kind of ``convolution product'' giving the
composition of two such functors. More precisely, given smooth algebraic varieties $X$, $Y$, $Z$, and
elements $E\in D(X\times Y)$ and $G\in D(Y\times Z)$ we can define $G\circ E\in D(X\times Z)$ by
$$
G\circ E=R\pi_{{XZ}_*}(L\pi^*_{XY}E\otimes L\pi^*_{YZ}G)
$$
where $\pi_{XY}$, $\pi_{YZ}$, $\pi_{XZ}$ are the projections from $X\times Y\times Z$ to the 
pairwise products giving a natural isomorphism of functors
$$
\bS^{G}\circ\bS^{E}=\bS^{G\circ E}.
$$
Another anology to the Fourier transform can be drawn. For this 
assume that we have sheaves $F$ and $G$ which only have one non-vanishing Fourier-Mukai transform, the $i$-th
one  ${ \Phi}  ^i({F})$ in the case of ${F}$ and the
$j$-th one ${ \Phi}  ^j(G)$ in the case of $G$. Given such sheaves there is the Parseval formula$$
\Ext_X^h({F},G)=\Ext_{\hat X}^{h+i-j}({ \Phi}^i({F}),{\Phi}^j(G))\,,
$$ 
which gives a correspondence between the extensions of $F$,
$G$ and the extensions of their Fourier-Mukai transforms. This formula can be considered as the analog of the Parseval formula for the ordinary Fourier transform for functions on a torus.

The Parseval formula can be proven using two facts. First, for 
arbitrary coherent sheaves $E$, $G$ the ext-groups can be
computed in terms of the derived category, namely
$$
\Ext^i(E,G)=\Hom_{D(X)}(E,G[i]).
$$
Secondly, the Fourier-Mukai transforms of
${F}$ and $G$ in the derived category $D(X)$ are given by
${\Phi}  ({F})={\Phi}^i({F})[-i]$ and ${\Phi}  (G)={\Phi} ^j(G)[-j]$.
Since the Fourier-Mukai transform is an equivalence of categories, we have
$$
\Hom_{D(X)}({F},G[i])=\Hom_{D(X)}({\Phi}  ^i({F}),{\Phi}^j(G)[i-j+h])
$$
implying the Parseval formula.

Some simple examples of geometrical and Fourier-Mukai functors can be given:
\begin{enumerate}
\item Let $X_{ell}$ be an elliptic curve with a distinguished point $p_0$ and consider as kernel 
the Poincar\'e line bundle ${\cal P}={\cal O}(D)$ where $D=\Delta-X_{ell}\times p_0-p_0\times X_{ell}$
(here $\Delta$ is the diagonal in $X_{ell}\times X_{ell}$). For $q$ a point on $X_{ell}$ we find that the
restriction of ${\cal P}$ to ${q\times X_{ell}}$ or $X_{ell}\times q$ is isomorphic to the line bundle ${\cal O}(q)\otimes {\cal O}(p_0)^{-1}$. As the even cohomology of $X_{ell}$ is spanned by $[X_{ell}]$ and the class of a point $[pt]$ and ${\rm Td}(X_{ell})=1$, we have ${\Phi_{\cal P}(k(p_0))={\cal O}_{X_{ell}}}$
and so $\Phi^{\nu({\cal P})}([pt])=[X_{ell}]$ (where $\Phi^{\nu(\cal P)}\colon H^*(X_{ell},{\bbQ})\to H^*(X_{ell},{\bbQ})$, see below). Moreover, using cohomology and base change we get
$h^1(\Phi_{\cal P}({\cal O}_{X_{ell}}))=R^1\pi_{1\ast}{\cal P}=k(p_0)$ (here $\pi_1$ is the projection on the first factor of ${X_{ell}}\times {X_{ell}}$). If we insert this into the base change 
theorem for $R^0\pi_{1\ast}{\cal P}$ and use the fact that $R^0(.)$ is torsion free, we find $h^0(\Phi_{\cal P})({\cal O}_{X_{ell}}))=k(p_0)[-1]$. It follows that the action of $\Phi^{\nu(\cal P)}$ on cohomology is given by ${\begin{pmatrix} 0 & 1\\ -1& 0\end{pmatrix}}$ using the basis $([{X_{ell}}],[pt])$ of $H^{\rm even}(X_{ell},{\bbQ})$.
\item
Let $F$ be the complex in $D(X\times X)$ defined by the 
the structure sheaf  $\Oc_{\Delta}$ of the diagonal $\Delta\subset X\times X$.
Then it is easy to check that $\bS^{F}\colon D(X)\to D(X)$ is isomorphic to the
identity functor on $D(X)$. 

If we shift degrees by $n$ taking $F=\Oc_{\Delta}[n]$ (a complex with only the sheaf $\Oc_\Delta$ placed in degree $n$), then $\bS^{F}\colon D(X)\to D(X)$ is the degree shifting functor $\G\mapsto \G[n]$.
\item
More generally,  given a proper morphism $f\colon X\to \X$, 
by taking as $F$ the structure sheaf of the
graph $\Gamma_f\subset X\times Y$, we have isomorphisms
of functors $\bS^{F}\simeq R f_\ast$ as functors  $D(X)\to D(\X)$ and 
$\bS^{F}\simeq f^\ast$ now as functors $D(\X)\to D(X)$.
\item 
Take $\X=X$ and let $L$ be a line bundle on $X$. If $F=\hat\pi^\ast L$, then $\bS^{F}(G)=G\otimes L$ for any $G$ in $D(X)$. 
\end{enumerate}

It is interesting to note that we cannot find examples of equivalences of derived categories other than \FMF s. This is due to a result by Orlov saying that if $X$ and $Y$ are smooth projective varieties then any fully faithful functor $D(X)\to D(Y)$ is a \gif. In particular, any equivalence of categories $D(X)\iso D(Y)$ is a \FMF.

Now it is often convenient to study problems for families rather than for single varieties. The main advantage of the relative setting is that base-change properties (or parameter dependencies) are better encoded into the problem. We can do that for \gif s as well.  To this end, we consider two morphisms $p\colon X\to B$, $\hat p\colon\X\to B$ of algebraic varieties.  We will assume that the morphisms are flat and so give nice families of algebraic varieties. We shall define a 
relative \gif\ in this setting by means of a ``kernel'' $E$ in the derived category $D(X\times_B\X)$, just by mimicking the ``abosulte'' definition. Moreover, we want \gif s for families and to allow further changes in the base space $B$, that is, we consider base-change morphisms $g\colon S\to B$. We denote all objects obtained by base
change to $S$ by a subscript $S$, like $X_S=S\fp X$ etc. In particular, the kernel $E$ defines an object $E_S=Lg^\ast E\in D((X\fp\X)_S)=D(X_S\times_S\X_S)$.

There is then a diagram
$$
\xymatrix{\save-<2.3truecm,0pt>*{(X\fp\X)_S\simeq }\restore{}
X_S\times_S\X_S \ar[d]^{\pi_S}\ar[r]^{\hat\pi_S}&\X_S
\ar[d]_{\hat p_S}\\ X_S \ar[r]^{p_S}& S}
$$ 
and the {relative \gif} associated to $E$ is the functor between the
derived  categories of quasi-coherent sheaves given by
$$
\bS^{E_S}\colon D(X_S)\to D(\X_S) \,,\quad F\mapsto \bS^{E_S}(F)=
R\hat\pi_{S\ast}(L\pi_S^\ast F\otimes E_S)
$$ 
(the tensor product is made in the derived category). When $\hat p$ is a flat morphism, $\pi_S$ is flat as well and we can simply write $\pi_S^\ast F$ instead of $L\pi_S^\ast F$.

We immediately note that the relative \gif\  with respect to $E\in D(X\times_B\X)$ {is nothing but the absolute \gif\  with kernel $i_\ast E\in D(X\times\X)$, where $i\colon X\fp\X\hookrightarrow X\times\X$ is the immersion}. The gain is that we can state neatly the following base-change property. For this we consider an object $F$ in $D(\X_S)$. It can then be shown that for every morphism $g\colon S'\to S$ there is an isomorphism
$$ Lg_{\X}^\ast(\bS^{E_S}(F))\simeq \bS^{E_{S'}}(L g_X^\ast F)
$$ in the derived category $D^-(\X_{S'})$,  where
$g_X\colon X_{S'}\to X_S$, $g_{\X}\colon\X_{S'}\to\X_S$ are the
morphisms induced by $g$.

Note also that the Parseval formula for the relative Fourier-Mukai transform has been
proved by Mukai in his original Fourier-Mukai transform for abelian
varieties, but can be easily extended to any situation in which a
Fourier-Mukai transform is an equivalence of categories.

For physical applications it is often convenient to work
in cohomology $H^*(X,{\bbQ})$. The passage from $D(X)$ to $H^*(X,{\bbQ})$ can be discribed as follows. We first send a complex $Z \in D(X)$ to its natural class in the K-group; we make then use of the fact that the Chern character $ch$ maps $K(X)\to CH^*(X)\otimes {\bbQ}$ and finally we apply the cycle map to $H^*(X,{\bbQ})$. This passage (by abuse of notation) is often denoted by $ch\colon D(X)\to H^*(X,{\bbQ})$, it commutes with pull-backs and transforms tensor products into dot products. Moreover, if we substitute the Mukai vector $v(Z )=ch(Z)\sqrt {Td(X)}$ for the Chern character $ch(Z)$ then we find the commutative diagramm
$$
\xymatrix{
D(X) \ar[d]_{v}\ar[r]^{\Phi^E}&D(Y)
\ar[d]_v\\ H^*(X,{\bbQ}) \ar[r]^{\Phi^{v(E)}}& H^*(Y,{\bbQ})}
$$ 
This can be shown using the Grothendieck-Riemann-Roch theorem and the fact that the power series 
defining the Todd class starts with constant term 1 and thus is invertible. 

\section{Elliptic fibrations and Weierstrass model}

An elliptic curve is a genus one Riemann surface and so can be
embedded in the two dimensional complex projective space ${\Ps}^2$. A simple embedding
is given by the homogeneous Weierstrass equation 
$$y^2z=4x^3-g_2xz^2-g_3z^3$$
where $x$, $y$ and $z$ are complex homogeneous coordinates on ${\Ps}^2$, thus
$(\lambda x,\lambda y, \lambda z)$ is identified with $(x,y, z)$ for any non-zero complex 
number $\lambda$. The parameters $g_2$ and $g_3$ are constants and encode the
different complex structures. If $z\neq 0$ we can rescale to affine coordinates where $z=1$.
Then it follows, viewed as a map from $x$ to $y$, that there are two branch cuts in the $x$-plane,
linking $x=\infty$ and the three roots of the cubic equation $4x^3-g_2x-g_3=0.$ The elliptic curve becomes singular if any two of these points coincide. The singular behaviour is characterized 
by the discriminant 
$$\Delta(g_2,g_3)=g_2^3-27g_3^2.$$
If $\Delta(g_2,g_3)\neq 0$, then the elliptic curve defined by the Weierstrass equation is a
smooth cubic curve, otherwise, if $\Delta(g_2,g_3)=0$, the curve is singular with arithmetic 
genus $\chi_a=1.$ Smooth points of the locus $\Delta(g_2,g_3)=0$ correspond to
rational curves with a single node. The point $g_2=g_3=0$ is the unique singular point
of the discriminant locus and corresponds to a rational curve with a single cusp. 

If the elliptic curve is considered as the complex plane modulo a discrete group of
translations, the group structure on the curve becomes transparent. Addition of points
in the complexe plane induces a natural notion of addition of points on the torus. 
If translated to the Weierstrass equation, the identity element corresponds to the point 
where ${x\over z}$ and ${y\over z}$ become infinite. The distinguished point $p$ on the 
elliptic curve, the zero in the group law, is given in affine coordinates by $x=y=\infty$ and can be scaled elsewhere in non-affine coordinates, such as to $x=z=0, y=1.$

We can now describe intuitively an elliptic fibration $X$ over a base manifold $B$ by giving the elliptic curve $E$ over each point in $B$. In the following it will be assumed that the fibration has a global section $\sigma.$ This requires that locally the parameters $g_2$ and $g_3$ are given as functions on the base. Globally, $g_2$ and $g_3$ will be sections of ${\cal L}^4$ and ${\cal L}^6$ with ${\cal L}$ 
being a line bundle on $B$ given as the conormal bundle $N_{B/X}^{-1}$ to the section $\sigma$ in $X$. From the equation of the discriminant it follows that $\Delta$ is a section of ${\cal L}^{12}$. The zeros of $\Delta$ define a divisor given by a number of points, a curve or a complex surface, in case
$B$ is complex one, two- or three-dimensional. 

More precisely, an elliptic fibration is a proper flat
morphism $\pi\colon X\to  B$ of schemes whose fibres are Gorenstein curves of arithmetic genus 1.  We also assume that $\pi$ has a {section} $\sigma\colon B\hookrightarrow X$ taking values in the smooth locus $X'\to B$ of $\pi$.  The generic fibres are then smooth elliptic curves whereas some singular fibres are allowed. If the base $B$ is a smooth curve, elliptic fibrations were studied and classified by Kodaira, who described all the types of singular fibres that may occur, the so-called Kodaira curves. When the base is a smooth surface, more complicated configuration of singular curves can occur and have been studied by Miranda.  

We denote by  $\sigma=\sigma(B)$  the image of the section, by $X_t$ the fibre of $\pi$ over $t\in B$ and by $i_t\colon X_t\hookrightarrow X$ the inclusion.
$\omega_{X/B}$  is the relative dualizing sheaf and we write $\omega=R^1\pi_\ast\Oc_X\iso (\pi_\ast\omega_{X/B})^\ast$, where the isomorphism is Grothendieck-Serre duality for $\pi$. The sheaf $L=\pi_\ast\omega_{X/B}$ is a line bundle  and
$\omega_{X/B}= \pi^\ast L$.  We write $\bar K=c_1(L)$. Adjunction formula for $\sigma\hookrightarrow X$ gives that
$\sigma^2=-\sigma\cdot \pi^{-1}\bar K$ as cycles on $X$.

Now if $B$ is a smooth curve, then from Kodaira's classification of possible singular fibres we find that the components of reducible fibres of $\pi$ which do not meet $\sigma $ form 
rational double point configurations disjoint from $\sigma$. Let $X\to {\bar{X}}$ be the result
of contracting these configurations and let $\bar{\pi}\colon {\bar{X}}\to B$ be the induced map.
Then all fibres of ${\bar \pi}$ are irreducible with at worst nodes or cusps as singularities. In this case 
we refer to ${\bar{X}}$ as the Weierstrass model of $X$.  The Weierstrass model can be constructed as follows:  the divisor $3\sigma$ is relatively ample and if $\Ec=\pi_\ast \Oc_X(3\sigma)=\Oc_B\oplus\omega^{\otimes 2}\oplus\omega^{\otimes3}$ and $\bar \pi\colon P=\Ps(\Ec^\ast)=\op{Proj}(S^\bullet(\Ec))\to B$ is the associated projective bundle, there is a projective morphism of $B$-schemes $j\colon X\to P$ such that $\bar X=j(X)$. Moreover, let $x:{\cal E}\to {\omega}^2$ be the natural projection, similar $y:{\cal E}\to {\omega}^3$ and $z:{\cal E}\to {\cal O}_B$ then $x, y, z$ are coordinates on ${\cal E}$. It follows that $x^3$ can be viewed as section of ${\rm Sym}^3{\cal E}^*\otimes {\omega}^6$ and similar for $y^2z$, $g_2xz^2$ and $g_3z^3$ and 
$s=y^2z-4x^3-g_2xz^2-g_3z^3$ is a global section of ${\rm Sym}^3{\cal E}^*\otimes {\omega}^6$.
Thus on each fibre ${\cal E}_t$ of ${\cal E}$, $s$ defines and element of ${\rm Sym}^3{\cal E}^*_t$
well defined up to a non-zero scalar. An element of ${\rm Sym}^3{\cal E}_t^*$ is a homogenous 
polynomial on ${\cal E}_t$ of degree 3, defining a cubic curve in $\Ps({\cal E}_t)$; and these
curves fit together to a family in $ {\Ps}({\cal E})$.  

By the sake of simplicity we shall refer to a particular kind of elliptic fibrations, namely elliptic fibrations with a section as above whose fibres are all {geometrically integral}. This means that the fibration is isomorphic with its Weierstrass model.
Now, special fibres can have at most one singular point, either a cusp or a simple node.  
Thus, in this case $3\sigma$ is relatively very ample and gives rise to a closed immersion
$j\colon X\hookrightarrow P$  such that $j^\ast\Oc_P(1)=\Oc_X(3\sigma)$.
Moreover $j$ is locally a complete intersection whose normal  sheaf is
$$
\Nc(X/P)\iso \pi^\ast\omega^{-\otimes 6}\otimes\Oc_X(9\sigma)\,.
$$
This follows by relative duality since
$$
\omega_{P/B}=\bigwedge \Omega_{P/B}\iso
\bar \pi^\ast\omega^{\otimes5}(-3)\,,
$$
 due to the Euler exact sequence 
$$
0\to\Omega_{P/B}\to \bar \pi^\ast \Ec(-1)\to\Oc_P\to 0\,.
$$
The morphism $\pi\colon X\to B$ is then a local complete intersection morphism (cf., for details 
see the book of Fulton) and has a virtual relative
tangent bundle $T_{X/B}=[j^\ast T_{P/B}]-[\Nc_{X/P}]$ in the $K$-group
$K^\bullet (X)$. The Todd class of the virtual tangent bundle $T_{X/B}$ is
$$
{\rm Td}(T_{X/B})=1-\tfrac12\, \pi^{-1}\bar K+ \frac{1}{12}(12\sigma\cdot 
\pi^{-1}\bar K+13\pi^{-1}\bar K^2)-\frac 12 \sigma\cdot \pi^{-1} \bar K^2+\text{ terms of higher degree.}
$$
This expression and the expression for the Todd class of the base manifold allow the computation of
the characteristic classes of the elliptic fibration. If we impose the Calabi-Yau condition to the elliptic
fibration (i.e., we set $\bar K=c_1(B)$ and so enforce $c_1(X)=0$) then the characteristic classes of $X$ 
are given by
\begin{align*}
 c_2(X)&=\pi ^*c_2(B)+11\pi^*c_1^2(B)+12\sigma \pi^*c_1(B) \\
 c_3(X)&=-60\pi^*c_1^2(B). \\
\end{align*}
When $X$ is a Calabi-Yau threefold, the presence of the section imposes constraints to the base surface $B$; it is known that it  has to be of a particular kind, namely $B$ has to be a del Pezzo surface (a surface whose anticanonical divisor $-K_B$ is ample), a Hirzebruch surface (a rational ruled surface), a Enriques surface (a minimal surface with $2 K_B$ numerically equivalent to zero) or a blow-up of a Hirzebruch surface. 

\section{Vector bundles for heterotic strings}

A compactification of the ten-dimen\-sional heterotic string is 
given by a holomorphic, stable $G$-bundle $V$ over a Calabi-Yau manifold $X$. The 
Calabi-Yau condition, the holomorphy and stability of $V$ are a direct consequence of the 
required supersymmetry in the uncompactified space-time. We assume that the underlying
ten-dimensional space $M_{10}$ is decomposed as $M_{10}=M_{4}\times X$ where $M_{4}$ 
(the uncompactified space-time) denotes the four-dimensional Minkowski space and $X$ a six-dimensional compact space
given by a Calabi-Yau threefold. To be more precise: supersymmetry 
requires that the connection $A$ on $V$ satisfies 
$$F_A^{2,0}=F_A^{0,2}=0, \qquad F^{1,1}\wedge J^2=0\,.$$
It follows that the connection has to be a holomorphic connection on a holomorphic 
vector bundle and in addition to satisfy the Donaldson-Uhlenbeck-Yau
equation that has a unique solution if the vector bundle is $\mu$-stable. 

In addition to $X$ and $V$ we have to specify a $B$-field on $X$ of field strength $H$. In order 
to get an anomaly free theory, the Lie group $G$ is fixed to be either $E_8\times E_8$ or 
$Spin(32)/{{\Z}_2}$ or one of their subgroups and $H$ has to satisfy the identity 
$$dH=\tr R\wedge R-\Tr F\wedge F$$ 
where $R$ and $F$ are the associated curvature forms of the spin connection on $X$ and the gauge connection on $V$. Also $\tr$ refers to the trace of the composite endomorphism of the tangent bundle to $X$ and $\Tr$ denotes the trace in the adjoint representation of $G$. For any closed four-dimensional submanifold $X_4$ of the ten-dimensional space-time $M_{10}$, the four form
$\tr R\wedge R-\Tr F\wedge F$ must have trivial cohomology.
Thus a necessary topological condition $V$ has to satisfy is $ch_2(TX)=ch_2(V)$
which simplifies to $c_2(TX)=c_2(V)$ for Calabi-Yau manifolds and $V$ being a $SU(n)$ vector bundle.

A physical interpretation of the third Chern-class can be given as a result of the decomposition of the ten-dimensional space-time into a four-dimensional flat Minkowski space and $X$. 
The decomposition of the corresponding ten-dimensional Dirac operator with values in $V$ shows that massless four-dimensional fermions are in one to one correspondence with zero modes of
the Dirac operator $D_V$ on $X$. The index of $D_V$ can be effectively computed using the
Hirzebruch-Riemann-Roch theorem and is given by
$${\rm index}(D)=\int_X \td (X)ch(V)={\frac{1}{ 2}}\int_X c_3(V)\,,$$
equivalently, we can write the index as ${\rm index}(D)=\sum_{i=0}^3(-1)^k \dim H^k(X,V)$.
For stable vector bundles we have $H^0(X,V)=H^3(X,V)=0$ and so the index 
computes the net-number of fermion generations $N_{\rm gen}$ in the respective
model.  

Now it has been observed that the inclusion of background five-branes changes the anomaly constraint . Various five-brane solutions 
of the heterotic string equations of motion have been discussed in: the gauge five-brane, the symmetric five-brane and the neutral five-brane. It has been shown that the gauge and symmetric five-brane solution involve finite size instantons of an unbroken
non-Abelian gauge group. In contrast, the neutral five-branes can be
interpreted as zero size instantons of the $SO(32)$ heterotic string.
The magnetic five-brane contributes a source term to the Bianchi identity for the three-form $H$,
$$dH=\tr R\wedge R-\Tr F\wedge F-n_5\sum_{\rm five-branes}\delta_5^{(4)}$$
and integration over a four-cycle in $X$ gives the anomaly constraint
$$c_2(TX)=c_2(V)+[W]\,.$$
The new term $\delta_5^{(4)}$ is a current that integrates to one in the direction 
transverse to a single five-brane whose class is denoted by $[W]$. The class $[W]$ is the Poincar\'e
dual of an integer sum of all these sources and thus $[W]$ should be a integral class, representing 
a class in $H_2(X,{\Z})$.  $[W]$ can be further specified taking into account
that supersymmetry requires that five-branes are wrapped on holomorphic curves thus $[W]$ 
must correspond to the homology class of holomorphic curves. This fact constraints $[W]$ to be
an algebraic class. Further, algebraic classes include negative classes, however, these lead to negative magnetic charges, which are un-physical, and so they have to be excluded. This constraints $[W]$ to be an effective class. Thus for a given Calabi-Yau threefold $X$ the effectivity of $[W]$ constraints the choice of vector bundles $V$.  

The study of the correspondence between the heterotic string (on an elliptic Calabi-Yau threefold) and F-theory (on an elliptic Calabi-Yau fourfold) has led Friedman, Morgan and Witten to introduce
a new class of vector bundles which satisfy the anomaly constraint with $[W]$ non-zero. As a result, they prove that the number obtained by integration of $[W]$ over the elliptic fibres of the Calabi-Yau threefold agrees with the number of three-branes given by the Euler characteristic of the Calabi-Yau fourfold divided by 24. Friedman, Morgan and Witten's study also shows that the Fourier-Mukai transform enters naturally into heterotic string theory.

We now review the strategy of Friedman, Morgan and Witten to elegantly construct
vector bundles on elliptic fibered Calabi-Yau manifolds. 
The bundle construction starts with a description of vector bundles
on an elliptic curve. A $SU(n)$ bundle on an elliptic curve $E$ is a rank $n$ vector bundle $V$ of trivial determinant. Any $SU(n)$ vector bundle $V$ on $E$ can be expressed, as a smooth bundle,
in the form $V=\oplus_{i=1}^n{L}_i$ where ${L}_i$ are holomorphic line bundles. Note that, as a holomorphic vector bundle, $V$ has only a filtration with quotients given by the ${L}_i$'s. The $SU(n)$ condition is imposed if $\otimes_{i=1}^n{L}_i$ is the trivial line bundle. Further, $V$ is semistable if all ${L}_i$ are of degree zero. The ${L}_i$ are uniquely determined up to permutations and further determine a unique point $q_i$ in $E$; conversely, every $q_i$ in $E$ determines a degree zero line
bundle ${L}_i={\Oc}(q_i-p)$ on $E$. Further, a semistable smooth $SU(n)$ bundle $V=\oplus_{i=1}^n{L}_i$ is determined due to the additional condition $\sum_{i=1}^n(q_i-p)=0$ (as divisors). The $q_i$ can be realized as roots of 
$$
s=a_0+a_2x+a_3y+a_4x^2+a_5x^2y+\dots+a_nx^{n/2}=0
$$
and give a moduli space of bundles ${\Ps}^{n-1}$, as stated by a theorem of Looijenga.
Note that if $n$ is odd, the last term in $s=0$ is given by $a_nx^{(n-3)/2}y$. The roots are determined by the coefficients $a_i$ only up to an overall scale factor so that the $a_i$ become the homogeneous coordinates on ${\Ps}^{n-1}$. 

The construction proceeds with a description of $SU(n)$ vector bundles on $X$. It is the basic idea to use first the bundle description of $SU(n)$ bundles on $E$ and then ``glue'' the bundle data together over the base manifold $B$ of $X$. More precisely, the variation of the $n$ points in the fibre over $B$ leads to a hypersurface
$C$ embedded in $X$, that is, $C$ is a ramified $n$-fold cover -the spectral cover- of the base. 
The line bundle on $X$ determined by $C$ is given by
${\Oc}_X(C)={\Oc}_X(n\sigma)\otimes {\M}$
where $\sigma$ denotes the section and ${\M}$ is a line bundle on $X$ whose restriction to the fibre is of degree zero. It follows that the cohomology class of $C$ in $H^2(X,{\Z})$ is given by 
$[C]=n\sigma +\eta,$
where $\eta=c_1({\M})$. 
Let $\pi_C\colon C\to  B$ and denote by $E_b$ the general elliptic fibre over a point $b$ in $B$, then we have $C\cap E_b=\pi_C^{-1}(b)=q_1+\ldots+q_n$ and $\sigma\cap E_b=p$. Now, each $q_i$ determines a line bundle ${L}_i$ of degree zero on $E_b$ whose sections are the meromorphic functions on $E_b$ with first order poles at $q_i$ and vanishing at $p$. The restriction of $V$ to $E_b$ is then $\rest V{E_b}=\oplus_{i=1}^n {L}_i$ as smooth bundles. As $b$ moves in the base the ${L}_i$ move in one to one correspondence with the $n$ points $q_i$ above $b$. This specifies a unique line bundle ${L}$ on $C$ such that $\pi_{C\ast}{L}=\rest V{B}$. 
Thus in addition to $C$ we have to specify a line bundle ${L}$ on $C$ to completely specify a rank $n$ vector bundle on $X$. 

This procedure leads typically to $U(n)$ vector bundles. To reduce to the $SU(n)$ case 
two further conditions have to be imposed: ${\M}$ if restricted to the fibre is the trivial bundle and ${L}$ has to be specified such that $c_1(V)=0$. 

We can now give a coherent description of $V$ using a relative Fourier-Mukai transform. For this we note that sheaves on an elliptic fibrations having degree zero on the fibres, admit a spectral cover description and can be reconstructed by the inverse Fourier-Mukai transformation from their spectral data. This is the reason why the condition of vanishing degree on the fibres is usually imposed, an assumption we will adopt here. 

For the description of the relative Fourier-Mukai transform it is appropriate to work on
$X\times_B {\tilde X}$ where $\tilde X$ is the compactified relative
Jacobian of $X$. $\tilde X$ parametrizes torsion-free rank 1 and
degree zero sheaves on the fibres of $X\to B$  and it is actually
isomorphic with $X$ so that we can identify
$\tilde X$ with $X$. We have a diagram:
$$
\xymatrix{
X\times_B{X} \ar[r]^{\pi_2} \ar[d]^{\pi_1}  & X \ar[d]^{\pi} \\
X\ar[r]^\pi & B
}
$$
and the corresponding kernel of the relative Fourier-Mukai transform is given by \emph{Poincar\'e} sheaf
$$
{\Pc}={\Oc}(\Delta)\otimes {\Oc}(- \pi_1^\ast \sigma)\otimes {\Oc}(-\pi_2^\ast \sigma)\otimes q^\ast K_B^{-1}
$$
normalized to make ${\Pc}$ trivial along $\sigma \times \tilde{X}$and
$X\times \sigma$. Here $q=\pi_1\circ \pi_1=\pi_2\circ \pi_2$ and $\Oc(\Delta)$ is the dual of the ideal sheaf of the diagonal, which is
torsion-free of rank 1.

In this way we can define two Fourier-Mukai transforms on the derived category $D(X)$ of bounded complexes of coherent sheaves on $X$ (we set $\hat{\Pc}={\Pc}^\ast \otimes q^\ast K_B^{-1}$).
$$
\begin{aligned}
{\Phi}({{\G}})&=R\pi_{1\ast }(\pi_2^\ast ({\G})\otimes {\Pc})\,,
\\ 
\hat {\Phi}({\G})&=R\pi_{1\ast }(\pi_2^\ast ({\G})\otimes \hat{\Pc})\,.
\end{aligned}
$$
We can also define the Fourier-Mukai functors ${\Phi}^i$ and
${\hat {\Phi} }^i$,
$i=0,1$ in terms of single sheaves $F$ by taking ${\Phi} ^i({F})$
and $\hat {\Phi}^i({F})$
as the $i$-th cohomology sheaves of the complexes ${\Phi}({F})$ and $\hat
{\Phi} ({F})$, we have
${\Phi}^i({F})=R^i\pi_{1\ast }(\pi_2^\ast ({F})\otimes {\Pc})\,$ (and $\hat {\Phi}  ^i({F})=R^i\pi_{1\ast }(\pi_2^\ast ({F})\otimes
\hat{\Pc})\,$).

Now it turns out that for the above description of vector bundles it is appropriate to take $F=i_\ast L$ with 
Chern characters given by ($\eta_E, {\eta}\in H^2(B,\bbQ)$)
\begin{align*}
\ch_0(i_\ast L)=0,\ 
\ch_1(i_\ast L)=n\sigma+\pi^\ast {\eta},\
\ch_2(i_\ast L)=\sigma \pi^\ast \eta_E+a_E F,\
\ch_3(i_\ast L)={s_E}.
\end{align*}
We can define a sheaf on $X$ as
$V={\Phi}^0(i_\ast L)$ (or alternatively as $\tilde V=\hat{\Phi}^0(i_\ast L)$) here $i\colon C\to X$ is the closed immersion of $C$ into $X$ ($\tilde V$ is related to $V$ by 
 $\tilde V=\tau^\ast V\otimes \pi^\ast \omega_B$). 
 
The characteristic classes of the rank $n$ vector bundle $V$ (and $\tilde V$) can be obtained if we apply the Grothendieck-Riemann-Roch theorem to the projection $\pi_1$ 
$$ch(V)=\pi_{1\ast}[\pi^\ast_2(i_\ast L)ch({\cal P})Td(T_{X/B})].$$ 
To make sure that the construction leads to $SU(n)$ vector bundles we set
$\eta_E=\frac12 nc_1$ giving $c_1(V)=0$
and the remaining Chern classes are given by
$$c_2(V)= \pi^\ast (\eta)\sigma+\pi^\ast (\varpi), \quad c_3(V)=-2\gamma_{\vert_{S}}$$
where
$$\varpi=\frac{1}{24}c_1(B)^2(n^3-n)+\frac{1}{2}(\lambda^2-\frac{1}{4})n\eta(\eta-nc_1(B)),$$ and
$\gamma\in H^{1,1}(C,{\Z})$ is some cohomology class satisfying $\pi_{C*}\gamma=0\in H^{1,1}(B,{\Z})$. The general solution for $\gamma$ has been derived by Friedman, Morgan and Witten and is given by
$\gamma=\lambda(n\sigma_{\vert_{C}}-\pi_C^\ast \eta+n\pi_C^\ast c_1(B))$ and $\gamma_{\vert_{S}}=-\lambda\pi^\ast \eta(\pi^\ast \eta-n\pi^\ast c_1(B))\sigma$$\gamma$ with $S=C\cap \sigma$. The parameter  $\lambda$ has to be determined such that $c_1(L)$ is an integer class. If $n$ is even, $\lambda=m$ ($m\in {\Z}$) and in addition we must impose $\eta=c_1(B)$ modulo 2. If $n$ is odd, $\lambda=m+\frac{1}{2}$.

It remains to discuss the stability of $V$. The stability depends on the properties of the defining data $C$ and $L$. If $C$ is irreducible and $L$ a line bundle over $C$ then $V$ will be a vector bundle stable with respect to
$$
J=\epsilon J_0+\pi^\ast H_B, \qquad {\epsilon} > 0
$$
if $\epsilon$ is sufficiently small. This has been proven by  Friedman, Morgan and Witten under the additional assumption that the restriction of $V$ to the generic fibre is regular and semistable. Here $J_0$ refers to some arbitrary K\"ahler class
on $X$ and $H_B$ a K\"{a}hler class on the base $B$. It implies that the bundle $V$ can be taken 
to be stable with respect to $J$ while keeping the volume
of the fibre $\mathfrak f$ of $X$ arbitrarily small compared to the volumes of effective curves
associated with the base. That $J$ is actually a good polarization can be seen by assuming 
$\epsilon=0$. Now we observe that  ${\pi}^\ast H_B$ is not a K\"{a}hler class on $X$ since
its integral is non-negative on each effective curve $C$ in $X$, however, there is one curve, the fibre $\mathfrak f$, where the integral vanishes. This means that ${\pi}^\ast H_B$ is on the boundary of
the K\"{a}hler cone and to make $V$ stable, we have to move slightly into the interior of the K\"{a}hler cone, that is, into the chamber which is closest to the boundary point ${\pi}^\ast H_B$. 
Also we note that although ${\pi}^\ast H_B$ is in the boundary of the K\"ahler cone, we can still define the slope $\mu_{{\pi}^\ast H_B}(V)$ with respect to it. Since $({\pi}^\ast H_B)^2$
is some positive multiple of the class of the fibre $\mathfrak f$, semi-stability with respect to ${\pi}^\ast H_B$ is implied by semi-stability of the restrictions $V {\vert}_\mathfrak f$ to the fibres.
Assume that $V$ is not stable with respect to $J$, then there is a
destabilizing sub-bundle $V' \subset V$ with $\mu_J(V') \ge \mu_J(V)$.
But semi-stability along the fibres says that $\mu_{{\pi}^\ast H_B}(V') \le
\mu_{{\pi}^\ast H_B}(V)$. If we had equality, it would follow that $V'$
arises by the spectral construction from a proper sub-variety of the
spectral cover of $V$, contradicting the assumption that this cover is
irreducible. So we must have a strict inequality $\mu_{{\pi}^\ast H_B}(V')
<\mu_{{\pi}^\ast H_B}(V)$. Now taking $\epsilon$ small enough, we can
ensure that $\mu_{J}(V') < \mu_{J}(V)$ thus $V'$ cannot destabilize $V$.

\section{D-branes and homological mirror symmetry}

Kontsevich proposed a homological mirror symmetry for a pair $(X,Y)$ of mirror dual Calabi-Yau manifolds; it is conjectured that there exists a categorical equivalence between the bounded derived 
category $D(X)$ and Fukaya's $A_{\infty}$ category ${\Fc}(Y)$ which is defined by using the symplectic structure on $Y$. An object of ${\Fc}(Y)$ is a special Lagrangian submanifolds with a flat $U(1)$ bundle on it. If we consider a family of manifolds $Y$ the object of ${\Fc}(Y)$ undergoes monodromy transformations. On the other side, the object of $D(X)$ is a complex of coherent sheaves on $X$ and under the categorical equivalence between $D(X)$ and ${\Fc}(Y)$ the monodromy (of three-cycles) is mapped to certain self-equivalences in $D(X)$. 

Since all elements in ${D}(X)$ can be represented by suitable
complexes of vector bundles on $X$, we can consider the topological 
K-group and the image $K_{\rm hol}(X)$ of ${D}(X)$. The Fourier-Mukai
transform $\bS^{\Ec}\colon {D}(X)\to {D}(X)$ induces then a corresponding 
automorphism $K_{\rm hol}(X)\to K_{\rm hol}(X)$ and also an automorphism
on $H^{\rm even}(X,{\bbQ})$ if we use the Chern character ring homomorphism
${\rm ch}:K(X)\to H^{\rm even}(X, {\bbQ})$, as described above. With this in mind we can introduce
various kernels and their associated monodromy transformations. 

For instance, let $D$ be a divisor in $X$ and consider the kernel ${\Oc}_{\Delta}(D)$
with $\Delta$ being the diagonal in $X\times X$; the corresponding Fourier-Mukai
transform acts on an object $G\in {D}(X)$ as twisting by a line bundle $G\otimes {\Oc}(D)$, this automorphism is then identified with the monodromy about the large complex
structure limit point (LCSL-point) in the complex structure moduli space.

Furthermore, if we consider the kernel given by the ideal sheaf  $\mathcal{I}_\Delta$ on $\Delta$ we
find that the action of $\bS^{\mathcal{I}_\Delta}$ on $H^{\rm even}(X)$ can be 
expressed by taking the Chern character ring homomorphism:
$$
\ch (\bS^{{\mathcal I}_\Delta}(G))=\ch_0(\bS^{\Oc_{X\times X}}(G))-\ch  (G)
=\left(\int \ch(G)\cdot \op\td (X)\right)-\ch(G)
$$

Kontsevich proposed that this automorphism should reproduce the monodromy
about the principal component of the discriminant of the mirror family $Y$. At the principal
component we have vanishing $S^3$ cycles (and the conifold singularity) thus this monodromy may
be identified with the Picard-Lefschetz formula.

Now for a given pair of mirror dual Calabi-Yau threefolds, it is generally assumed that $A$-type and $B$-type D-branes exchange under mirror symmetry. For such a pair Kontsevich's correspondence between automorphisms of $D(X)$ and monodromies of three-cycles can then be tested. More specifically, a comparison relies on the identification of two central charges associated to D-brane
configurations on both sides of the mirror pair. 

For this we have first  to specify a basis for the three-cycles $\Sigma_i\in H^3(Y,{\Z})$ such that the intersection form takes the canonical form $\Sigma_i\cdot \Sigma_j=\delta_{j,i+b_{2,1}+1}=\eta_{i,j}$ for $i=0,...,b_{2,1}$. It follows that a three-brane wrapped about the cycle $\Sigma=\sum_i n_i\Sigma^i$ has an (electric,magnetic) charge vector ${\bf n}=(n_i)$. The periods of the holomorphic three-form $\Omega$
are then given by
$$
\Pi_i=\int_{\Sigma_i}\Omega
$$
and can be used to provide projective coordinates on the complex structure moduli space.
If we choose a symplectic basis $(A_i,B_j)$ of $H_2(Y,{\Z})$ then the 
$A_i$ periods serve as projective coordinates and the $B_j$ periods satisfy 
the relations $\Pi^j=\eta_{i,j}\partial {\cal F}/\partial \Pi^i$, where ${\cal F}$ is the prepotential which has near the large radius limit the asymptotic form (as analyzed by Candelas, Klemm, Theisen, Yau and Hosono, cf. further reading).
$$
{\cal F}=\frac 16 \sum_{abc}k_{abc}t_at_bt_c+\frac 12 \sum_{ab}c_{ab}t_at_b-
\sum_a \frac {c_2(X)J_a}{24} t_a+ \frac{\zeta(3)}{2(2\pi i)^3} \chi(X)+{\Oc}(q),
$$
where $\chi(X)$ is the Euler characteristic of $X$, $c_{ab}$ are rational constants (with $c_{ab}=c_{ba}$) reflecting an $Sp(2h^{11}+2)$ ambiguity and $k_{abc}$ is the classical triple intersection number given by
$$
k_{abc}=\int_X J_a\wedge J_b\wedge J_c.
$$
The periods determine the central charge $Z({\bf n})$ of a three-brane wrapped about the cycle $\Sigma=\sum_in_i[\Sigma_i]$ 
$$
Z({\mathbf n})=\int_{\Sigma}\Omega=\sum_i n_i\Pi_i,
$$
where $\sum_in_i\Pi_i=n_6\Pi_1+n_4^1\Pi_2+n_4^2\Pi_3+n_0\Pi_4+n_2^1\Pi_5+n_2^2\Pi_6$. The $\Pi$'s are given by the associated period vector
$$
\begin{pmatrix} \Pi_1\\ \Pi_2\\ \Pi_3 \\ \Pi_4 \\ \Pi_a \end{pmatrix}=
\begin{pmatrix}  \frac16 k_{abc}t_at_bt_c+\frac{c_2(X)J_a}{24}t_a\\
-\frac12 k_{abc}+c_{ab}t_b+\frac{c_2(X)J_a}{24}\\
1\\ t_a \end{pmatrix} \,.
$$

On the other side, the central charge associated with an object $E$ of $D(X)$ 
is given by 
$$
Z(E)=-\int_X e^{-t_aJ_a} \ch(E) (1+\frac{c_2(X)}{24}).
$$
The two central charges are to be identified under mirror symmetry. If we compare the two central charges $Z({\mathbf n})$ and $Z(E)$ then we obtain a map relating the Chern-classes of $E$ to the D-brane charges ${\mathbf n}$, we find
\begin{align*}
\ch_0(E)&=n_6\\
\ch_1(E)&=n_4^aJ_a\\
\ch_2(E)&=n_2^b+c_{ab}n_4^a\\
\ch_3(E)&=-n_0-\frac{c_2(X)J_b}{12}n_4^b.
\end{align*}

Insertion of the expressions for $\ch(E)$ in $\ch (\bS^{{\mathcal I}_\Delta}(E))$ yields a linear tranformation acting on ${\mathbf n}$
$$n_6\to n_6+n_0
$$
which agrees with the monodromy transformation about the conifold locus.

Similarly we find that the monodromy transformation about the large complex structure limit point
corresponds to automorphisms 
$$[E]\to [E\otimes {\Oc}_X(D)]
$$
where $D$ is the associated divisor defining the large radius limit in the K\"ahler moduli space.
Using the central charge identification, the automorphism/monodromy correspondence has been 
made explicit 
 for various dual pairs of mirror Calabi-Yau threefolds (given as hypersurfaces in weighted projective spaces).  This identification provides evidences for Kontsevich's proposal of homological mirror symmetry. 

\newpage
\subsection*{Further Reading}
{\sc C.~ Bartocci, U.~Bruzzo, D.~Hern{\'a}ndez Ruip{\'e}rez, and M.~Jardim},
{\em Nahm and {F}ourier-{M}ukai transforms in geometry and mathematical physics},
{To appear in Progress in Mathematical Physics, Birkha\"user, 2005.}\\
 [3ex]
{\sc A.~I. Bondal and D.~O. Orlov}, {\em Reconstruction of a variety from the
  derived category and groups of autoequivalences}, Compositio Math., 125
  (2001), pp.~327--344.\\
 [3ex]
{\sc C.~G. Callan, Jr., J.~A. Harvey, and A.~Strominger}, {\em Worldbrane
  actions for string solitons}, Nuclear Phys. B, 367 (1991), pp.~60--82.\\
[3ex]
{\sc P.~Candelas, A.~Font, S.~Katz, and D.~R. Morrison}, {\em Mirror symmetry
  for two-parameter models. {II}}, Nuclear Phys. B, 429 (1994), pp.~626--674.\\
[3ex]
{\sc R. Y. Donagi}, {\em Principal bundles on elliptic fibrations}, Asian J. Math. {\bf 1} (1997) 214-223,
alg-geom/9702002.\\
[3ex]
{\sc R. Y. Donagi}, {\em Taniguchi lecture on principal bundles on elliptic fibrations}, hep-th/98020094.\\
[3ex]
{\sc R.~Friedman, J.~W. Morgan, and E.~Witten}, {\em Vector bundles and {${\rm
  F}$} theory}, Comm. Math. Phys., 187 (1997), pp.~679--743.\\
[3ex]
{\sc W.~Fulton}, {\em Intersection theory}, vol.~2 of Ergebnisse der Mathematik
  und ihrer Grenzgebiete (3) [Results in Mathematics and Related Areas (3)],
  Springer-Verlag, Berlin, 1984.\\
[3ex]
{\sc S.~Hosono, A.~Klemm, S.~Theisen, and S.-T. Yau}, {\em Mirror symmetry,
  mirror map and applications to complete intersection {C}alabi-{Y}au spaces},
  Nuclear Phys. B, 433 (1995), pp.~501--552.\\
[3ex]
{\sc S.~Hosono}, {\em G{KZ} systems, {G}r\"obner fans, and moduli spaces of
  {C}alabi-{Y}au hypersurfaces}, in Topological field theory, primitive forms
  and related topics (Kyoto, 1996), vol.~160 of Progr. Math., Birkh\"auser
  Boston, Boston, MA, 1998, pp.~239--265.\\
[3ex]
{\sc Y.~Kawamata}, {\em {$D$}-equivalence and {$K$}-equivalence}, J.
  Differential Geom., 61 (2002), pp.~147--171.\\
[3ex]
{\sc K.~Kodaira}, {\em On compact analytic surfaces. {II}, {III}}, Ann. of
  Math. (2) 77 (1963), 563--626; ibid., 78 (1963), pp.~1--40.\\
[3ex]
{\sc M.~Kontsevich}, {\em Homological algebra of mirror symmetry}, in
  Proceedings of the International Congress of Mathematicians, Vol.\ 1, 2
  (Z\"urich, 1994), Basel, 1995, Birkh\"auser, pp.~120--139.\\
 [3ex]
{\sc R.~Miranda}, {\em Smooth models for elliptic threefolds}, in The
  birational geometry of degenerations (Cambridge, Mass., 1981), vol.~29 of
  Progr. Math., Birkh\"auser Boston, Mass., 1983, pp.~85--133.\\
 [3ex]
{\sc S.~Mukai}, {\em Duality between {$D(X)$} and {$D(\hat X)$} with its
  application to {P}icard sheaves}, Nagoya Math. J., 81 (1981), pp.~153--175.\\
[3ex]
{\sc D.~O. Orlov}, {\em Equivalences of derived categories and ${K}3$
  surfaces}, J. Math. Sci. (New York), 84 (1997), pp.~1361--1381.
\newblock Algebraic geometry, 7.\\
[3ex]
{\sc E.~Witten}, {\em New issues in manifolds of {${\rm SU}(3)$} holonomy},
  Nuclear Phys. B, 268 (1986), pp.~79--112.\\

\newpage
\subsection*{See also}
Derived categories, String dualities, Calabi-Yau manifolds, Mirror symmetry %

\newpage
\subsection*{Key words}
Derived categories\\
Fukaya category\\
Index theorems\\
Characteristic-classes\\
String dualities\\
Heterotic string theory\\
F-theory\\
Supersymmetry\\
D-branes\\
Calabi-Yau manifolds\\
Mirror symmetry\\

\end{document}